\documentclass{ws-procs975x65}

\begin{document}

\title{Effect of vector--axial-vector mixing
to dilepton spectrum\\
in hot and/or dense matter~\footnote{
Talk given by M.~Harada at 2009 Nagoya Global COE Workshop
``Strong Coupling Gauge Theories in LHC Era (SCGT09)''.
This talk is based on the work done in 
Refs.~\citen{our,VAmix-dense}
}
}

\author{Masayasu Harada}

\address{Department of Physics, Nagoya University, 
Nagoya, 464-8602, JAPAN\\
E-mail: harada@hken.phys.nagoya-u.ac.jp}

\author{Chihiro Sasaki}

\address{
Frankfurt Institute for Advanced Studies, 
J.W. Goethe University, 
D-60438 Frankfurt, Germany\\
Physik-Department,
Technische Universit\"{a}t M\"{u}nchen,
D-85747 Garching, Germany\\
E-mail: sasaki@fias.uni-frankfurt.de}

\begin{abstract}
In this write-up we summarize main results of 
our recent analyses on the mixing 
between transverse $\rho$ and $a_1$ mesons in hot and/or dense
matter.
We show that
the axial-vector meson contributes significantly 
to the vector spectral function in hot matter through the mixing.
In dense baryonic matter, we include a mixing
through a set of $\omega\rho a_1$-type interactions.
We show that a clear enhancement of the
vector spectral function appears below $\sqrt{s}=m_\rho$
for small three-momenta of the $\rho$ meson, 
and thus the vector spectrum exhibits broadening.
\end{abstract}


\bodymatter

\section{Introduction}

In-medium modifications of hadrons have been extensively
explored in the context of chiral dynamics of QCD~\cite{review,rapp}.
Due to an interaction with pions in the heat bath, the vector 
and axial-vector current correlators are mixed.
At low temperatures or densities a low-energy theorem based on 
chiral symmetry describes this V-A mixing~\cite{theorem}.
The effects to the thermal vector spectral function have been
studied through the theorem~\cite{vamix}, or using chiral reduction
formulas based on a virial expansion~\cite{chreduction}.

In Ref.~\citen{our}, 
it was shown that
the effects of the V-A mixing, and how the 
axial-vector mesons affect the spectral function near the chiral 
phase transition, within an effective field theory.
The analysis was carried out assuming several possible patterns 
of chiral symmetry restoration: dropping or non-dropping $\rho$
meson mass along with changing $a_1$ meson mass, 
both considered to be options from a
phenomenological point of view.

In Ref.~\citen{VAmix-dense},
we studied a novel effect of the V-A mixing 
through a set of $\omega\rho a_1$-type interactions
at finite baryon density, which was introduced by a
Chern-Simons term in a holographic QCD model~\cite{hqcd}.
We focused on the V-A mixing at tree level and its consequence
on
the in-medium spectral functions which are the main input 
to the experimental observables.
We showed that 
the mixing produces
a clear enhancement of the vector spectral function 
below $\sqrt{s}=m_\rho$,
and that
the vector spectral function 
is broadened due to the mixing.
We also discussed its relevance to dilepton measurements.

In this write-up, we summarize main results of these papers
especially focusing on the effect to the vector
spectral function.

\section{Effects of Vector--Axial-vector Mixing in Hot Matter}
\label{sec:hot}

\begin{figure}
\psfig{file=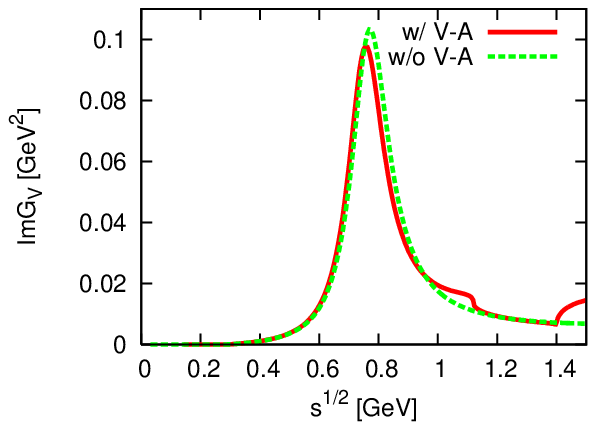,width=4cm}
\
\psfig{file=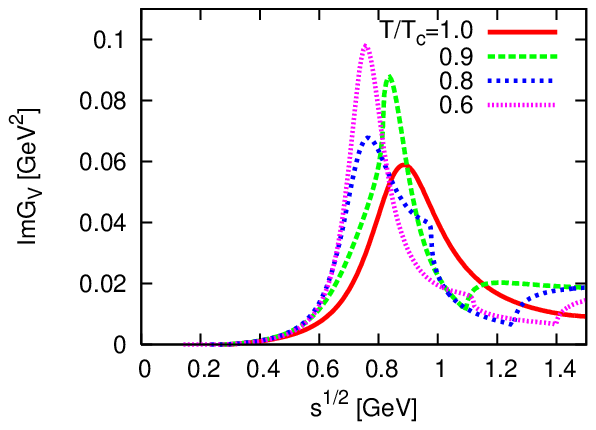,width=4cm}
\
\psfig{file=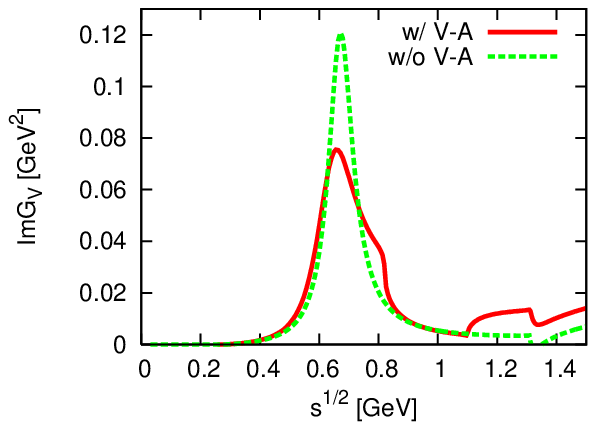,width=4cm}
\caption{Left figure shows the vector spectral function
at temperature $T/T_c = 0.6$. The solid (red) curve includes 
the effect of V-A mixing, while the dashed (green) curve does not.
The middle figure shows the spectral function (option (A))
for $m_\pi = 140$\,MeV at several temperatures $T/T_c = 0.6$-$1.0$.
The right figure shows the
vector spectral function (option (B))
for $m_\pi = 140$ MeV in type (I)
at temperature $T/T_c = 0.8$.
}\label{vamix}
\end{figure}

We start with showing
the vector spectral function at $T/T_c=0.6$,
without any dropping masses,
in Fig.~\ref{vamix} (left).
Two cases are compared; one includes the V-A mixing and the other
does not. Both the spectral functions has a peak at $M_\rho$. 
The effects of V-A mixing can be seen as
a shoulder at $\sqrt{s}=M_{a_1}- m_\pi$
and a bump above $\sqrt{s}=M_{a_1} + m_\pi$.

In Ref.~\citen{our},
two possible cases of chiral symmetry restoration are
studied:
\begin{description}
\item[(A)] Dropping $a_1$ meson mass but non-dropping $\rho$ mass;
\item[(B)] Dropping $\rho$ and $a_1$ meson masses.
\end{description}
In both cases, the ``flash temperature''~\cite{BLR} $T_f$
is introduced for controling how the mesons experience 
partial restoration of chiral symmetry.
Then, the masses of vector and/or axial-vector mesons are
assumed to have temperature dependences only above $T_f$.
In option (A) at the chiral limit, 
the dropping $a_1$ meson mass for $T > T_f$ is taken as
\begin{equation}
\frac{ M_{a_1}^2(T) - M_\rho^2 }{ M_{a_1}^2(T=0) - M_\rho^2 }
= \frac{ T_c^2 - T^2 }{ T_c^2 - T_f^2 }
\ ,
\end{equation}
where $T_c$ is the critical temperature of the chiral symmetry
restoration.
The vector spectral function for this option (A), with the
effect of pion mass included, is shown at several temperatures
in Fig.~\ref{vamix} (middle).
Below $T_c$ one observes
the previously mentioned threshold effects moving downward with 
increasing temperature. It is remarkable
that at $T_c$ the spectrum shows almost no traces of 
$a_1$-$\rho$-$\pi$ threshold effects:
The $\rho$ to $a_1$ mass ratio becomes almost $1$ at $T=T_c$
even though the effect of $M_\pi$ is included.
Furthermore,
one can show that
$a_1$-$\rho$-$\pi$ coupling constant becomes very tiny,
$g_{a_1 \rho\pi} \sim 0.06\, m_\pi$.
This
indicates that at $T_c$ {\it the $a_1$ meson mass nearly equals
the $\rho$ meson mass and the $a_1$-$\rho$-$\pi$ coupling
almost vanishes even in the presence of explicit
chiral symmetry breaking.}

In option (B), 
two types of temperature dependences were used in Ref.~\citen{our}:
\begin{eqnarray}
\mbox{(I)}&:&
 \frac{M_\rho^2(T)}{M_\rho^2(T=0)}
 = \frac{ T_c^2 - T^2 }{ T_c^2 - T_f^2 }
\ ,
\quad
 \frac{M_{a_1}^2(T) - M_\rho^2(T)}{M_{a_1}^2(T=0) - M_\rho^2(T=0)}
 = \left( \frac{ T_c^2 - T^2 }{ T_c^2 - T_f^2 } \right)^2
\ ,
\\
\mbox{(II)}&:&
 \frac{M_\rho^2(T)}{M_\rho^2(T=0)}
 = \left( \frac{ T_c^2 - T^2 }{ T_c^2 - T_f^2 } \right)^2
\ ,
\quad
 \frac{M_{a_1}^2(T) - M_\rho^2(T)}{M_{a_1}^2(T=0) - M_\rho^2(T=0)}
 = \frac{ T_c^2 - T^2 }{ T_c^2 - T_f^2 }
\ .
\end{eqnarray}
Figure~\ref{vamix} (right) shows 
the vector spectrum using the type (I) 
parameterization at $T = 0.8\,T_c$. 
The feature that the $a_1$ meson suppresses the vector spectral
function 
through the V-A mixing remains unchanged.
Compared with the curve for $T/T_c=0.8$ in Fig.~\ref{vamix} (middle),
a bump through the V-A mixing and the $\rho$ peak are shifted downward 
since both the $\rho$ and $a_1$ masses drop.
The self-energy has a cusp at the threshold $2\,M_\rho$ and
this appears as a dip at $\sqrt{s} \sim 1.3$ GeV.
The influence of finite $m_\pi$ turns out to be in threshold
effects as before.
We find that the nearly 
vanishing V-A mixing as seen for the non-dropping $\rho$ mass,
option (A).

The result given here shows
that the
axial-vector meson contributes significantly 
to the vector spectral function;
the presence of the $a_1$ 
reduces the vector spectrum around $M_\rho$
and enhances it around $M_{a_1}$.
A major change with both dropping $\rho$ and $a_1$ masses is
a systematic downward shift of the vector spectrum.
We observe that
the $a_1$-$\rho$-$\pi$ coupling almost 
vanishes at the critical temperature
$T_c$ and thus the V-A mixing becomes very tiny.

\section{Effects of Vector--Axial-vector Mixing in Dense Matter}
\label{sec:dense}

In this section,
we summarize main points shown in Ref.~\citen{VAmix-dense}.

At finite baryon density a system preserves parity 
but violates charge conjugation invariance.
Chiral Lagrangians thus in general build 
in the term
\begin{equation}
{\cal L}_{\rho a_1} = 2C\,\epsilon^{0\nu\lambda\sigma}
\mbox{tr}\left[ \partial_\nu V_\lambda \cdot A_\sigma
{}+ \partial_\nu A_\lambda \cdot V_\sigma \right]\,,
\label{vaterm}
\end{equation}
for the vector $V^\mu$ and axial-vector $A^\mu$ mesons with
the total anti-symmetric tensor $\epsilon^{0123}=1$ and a 
parameter $C$.
This mixing results in the dispersion relation~\cite{hqcd}
\begin{equation}
p_0^2 - \bar{p}^2 = \frac{1}{2}\left[ 
m_\rho^2 + m_{a_1}^2 \pm \sqrt{(m_{a_1}^2 - m_\rho^2)^2
{}+ 16 C^2 \bar{p}^2}
\right]\,,
\label{disp}
\end{equation}
which describes the propagation of a mixture of the transverse $\rho$ 
and $a_1$ mesons with non-vanishing three-momentum $|\vec{p}|=\bar{p}$.
The longitudinal polarizations, on the other hand,
follow the standard dispersion
relation, $p_0^2 - \bar{p}^2 = m_{\rho,a_1}^2$.
When the mixing vanishes as $\bar{p} \to 0$, Eq.~(\ref{disp}) with
lower sign provides $p_0 = m_\rho$ and it with upper sign
does $p_0 = m_{a_1}$.  In the following, we call the mode following
the dispersion relation with the lower sign in Eq.~(\ref{disp})
``the $\rho$ meson'', 
and that with the upper sign ``the $a_1$ meson''.

The mixing strength $C$ in Eq.~(\ref{vaterm}) can be 
estimated assuming the $\omega$-dominance
in the following way:
The gauged Wess-Zumino-Witten terms in 
an effective chiral Lagrangian include
the $\omega$-$\rho$-$a_1$ term~\cite{kaiser} which leads to
the following mixing term
\begin{equation}
{\cal L}_{\omega\rho a_1} = g_{\omega\rho a_1}
\langle \omega_0\rangle \epsilon^{0\nu\lambda\sigma}
\mbox{tr}\left[ \partial_\nu V_\lambda \cdot A_\sigma
{}+ \partial_\nu A_\lambda \cdot V_\sigma \right]\,,
\end{equation}
where the $\omega$ field is replaced with its expectation value
given by $\langle \omega_0\rangle = g_{\omega NN}\cdot n_B/m_\omega^2$.
One finds with empirical numbers
$C = g_{\omega\rho a_1}\langle \omega_0\rangle \simeq 0.1$\,GeV
at normal nuclear matter density. 
As we will show below, this is too small to have an importance
in the correlation functions.
In a holographic QCD approach, on the other hand,
the effects from an infinite tower of the $\omega$-type vector
mesons are summed up to give
$C \simeq 1\,\mbox{GeV}\cdot(n_B/n_0)$ 
with normal nuclear matter density $n_0 = 0.16$ fm$^{-3}$~\cite{hqcd}.
In the following we
assume an actual value of $C$ in QCD
in the range $0.1 < C < 1$\,GeV.
Some importance of the higher Kaluza-Klein
(KK) modes {\it even in vacuum} in the 
context of holographic QCD can be seen in the pion electromagnetic 
form factor at the photon on-shell:
This is saturated by the lowest four vector mesons in a top-down 
holographic QCD model~\cite{ss,matsuzaki2}. 
In hot and dense environment those higher members get modified
and the masses might be somewhat decreasing evidenced in an
in-medium holographic model~\cite{sstem}. This might provide
a strong V-A mixing $C > 0.1$\,GeV in three-color QCD
and the dilepton measurements may give a good testing ground.

\vspace{0.3cm}

\noindent
\begin{minipage}{6cm}
\ \ 
In Fig.~2,
we show the dispersion 
relations~(\ref{disp}) for the transverse modes together
with those for the longitudinal modes with $C = 1$  and
$0.5$\,GeV.
This shows that, when $C=0.5$\,GeV, 
there are only small changes for both $\rho$ and $a_1$ mesons,
while a substantial change for $\rho$ meson when $C=1$\,GeV.
For very large $\bar{p}$ the longitudinal and transverse 
dispersions are in parallel with a finite gap, $\pm C$. 
\ \par
\ 
\end{minipage}
\ \ 
\begin{minipage}{6cm}
\addtocounter{figure}{1}
\begin{flushleft}
\psfig{file=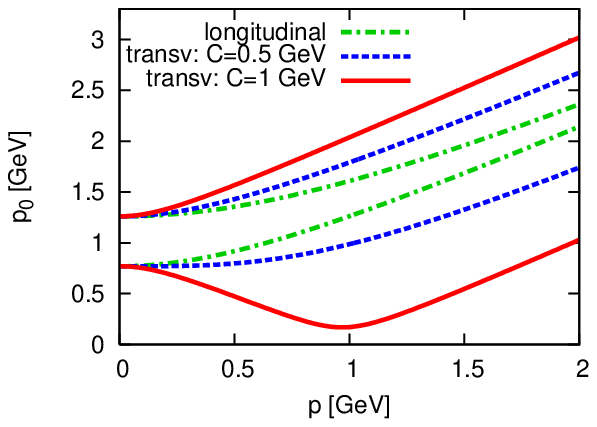,width=5cm}
{\small
\par
Fig.~2. The dispersion relations for the $\rho$ (lower 3 curves)
and $a_1$ (upper 3 curves) mesons for $C=0.5$, and $1$\,GeV.
\par
\ 
}
\end{flushleft}
\end{minipage}

\begin{figure}[htbp]
\begin{center}
\psfig{file=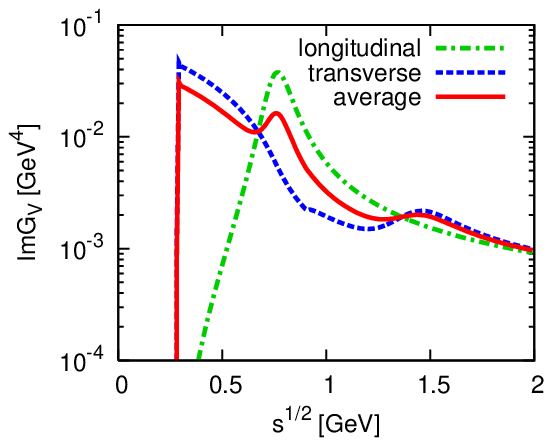,width=6cm}
\ 
\psfig{file=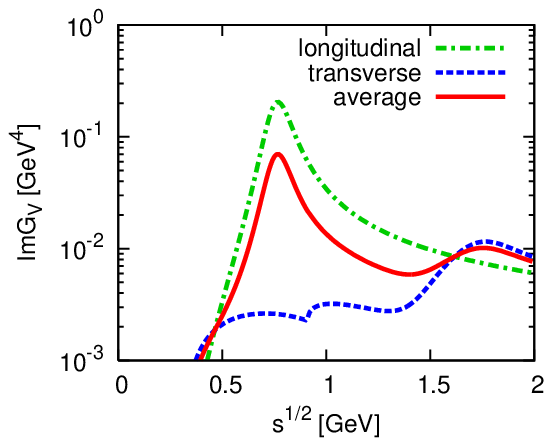,width=6cm}
\caption{
The vector spectral function for $C=1$\,GeV.
The curves of the left figure are calculated integrating over 
$0 < \bar{p} < 0.5$\,GeV, and those of the right figure
over $0.5 < \bar{p} < 1$\,GeV. 
Here we use the values of masses given by
$m_\pi = 0.14$\,GeV, $m_\rho = 0.77$\,GeV, 
$m_{a_1} = 1.26$\,GeV,
and the widths 
given by the imaginary 
part of one-loop diagrams in a chiral Lagrangian approach 
as~\cite{hs:ghls,VAmix-dense}
with the on-shell values of 
$\Gamma_{\rho}(s=m_\rho^2) = 0.15$\,GeV and
$\Gamma_{a_1}(s=m_{a_1}^2) = 0.33$\,GeV.
}
\label{CC1}
\end{center}
\end{figure}
In Fig.~\ref{CC1},
we plot the integrated spectrum over three momentum,
which is a main ingredient in dilepton production rates.
Figure~\ref{CC1} (left) shows a clear enhancement of the spectrum 
below $\sqrt{s}=m_\rho$ due to the mixing.
This enhancement becomes much 
suppressed when the $\rho$ meson is moving with a large three-momentum
as shown in Fig.~\ref{CC1} (right). The upper
bump now emerges more remarkably and becomes a clear indication
of the in-medium effect from the $a_1$ via the mixing.

As an application of the above in-medium spectrum, we calculate
the production rate of a lepton pair emitted from dense 
matter through a decaying virtual photon.
\begin{figure}[htbb]
\begin{center}
\psfig{file=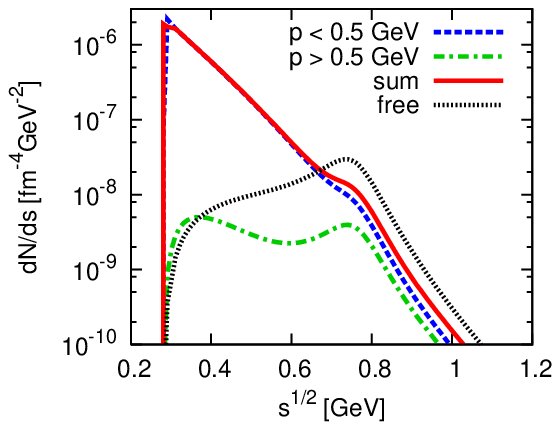,width=6cm}
\ 
\psfig{file=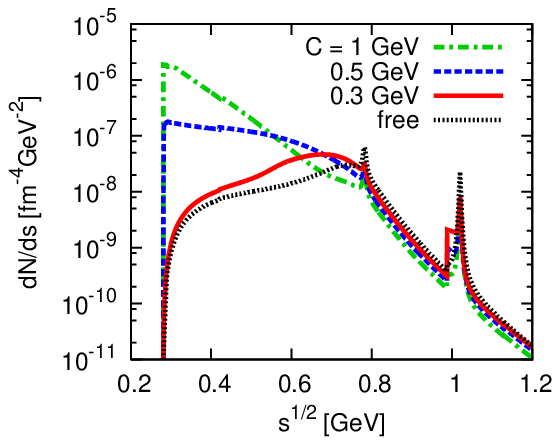,width=6cm}
\caption{
Left figure shows
the dilepton production rate at $T=0.1$\,GeV for $C=1$\,GeV. 
Integration over $0 < \bar{p} < 0.5$\,GeV (dashed) and 
$0.5 < \bar{p} < 1$\,GeV (dashed-dotted) was carried out.
The right figure shows
the dilepton production rate at $T=0.1$\,GeV
with various mixing strength $C$.
Integration over $0 < \bar{p} < 1$\,GeV was done.
We use the constant widths with values of
$\Gamma_\omega = 8.49$\,MeV, $\Gamma_\phi = 4.26$\,MeV,
$\Gamma_{f_1(1285)}=24.3$\,MeV and 
$\Gamma_{f_1(1420)}=54.9$\,MeV.
}
\label{rate}
\end{center}
\end{figure}
Figure~\ref{rate} (left)
presents the integrated rate at $T=0.1$\,GeV
for $C=1$\,GeV.
One clearly observes a strong three-momentum dependence and an 
enhancement below $\sqrt{s}=m_\rho$ due to the Bose
distribution function which result in a strong spectral broadening.
The total rate is mostly governed by the spectrum with 
low momenta $\bar{p}<0.5$\,GeV due to the large mixing parameter $C$.
When density is decreased, 
the mixing effect gets irrelevant and
consequently in-medium effect in low $\sqrt{s}$ region is reduced 
in compared with that at higher density.  
The calculation performed in hadronic many-body theory in fact
shows that the $\rho$ spectral function with a low momentum
carries details of medium modifications~\cite{riek}.
One may have a chance to observe it in heavy-ion collisions 
with certain low-momentum binning at J-PARC, GSI/FAIR and RHIC 
low-energy running.

It is straightforward to introduce other V-A mixing between 
$\omega$-$f_1(1285)$ and $\phi$-$f_1(1420)$.
In Fig.~\ref{rate} (right) 
we plot the integrated rate at $T=0.1$\,GeV
with several mixing strength $C$ which are phenomenological option.
One observes that the enhancement below $m_\rho$ is suppressed
with decreasing mixing strength. This forms into a broad bump
in low $\sqrt{s}$ region and its maximum moves toward $m_\rho$.
Similarly, some contributions are seen just below $m_\phi$.
This effect starts at threshold $\sqrt{s}= 2 m_K$. 
Self-consistent calculations of the spectrum in dense medium
will provide a smooth change and this eventually makes the $\phi$
meson peak somewhat broadened.

Finally, we remark that
the importance of the mixing effect studied here
relies on the coupling
strength $C$. 
Holographic QCD predicts an extremely strong 
mixing $C \sim 1$\,GeV at $n_B = n_0$ which leads to
vector meson condensation at $n_B \sim 1.1\,n_0$~\cite{hqcd}.
This may be excluded by known properties of nuclear
matter and therefore in reality the strength $C$ will be
smaller. We have discussed a possible range of $C$ to be
$0.1 < C < 1$\,GeV based on higher excitations and their in-medium 
modifications.
The parameter $C$ does carry an unknown density dependence.
This will be determined in an elaborated treatment of
hadronic matter along with the underlying QCD dynamics.
If $C \sim 0.1$\,GeV at $n_0$ were preferred as
the lowest-omega dominance, the mixing effect is irrelevant there.
However, it becomes more important at higher densities,
e.g. $C = 0.3$\,GeV at $n_B/n_0 = 3$ which leads to
a distinct modification from the spectrum in free space.

\section{Summary}

In this write-up,
we summarized main results of 
our recent analyses on the mixing 
between $\rho$ and $a_1$ mesons in hot and/or dense
matter.

The analysis in Ref.~\citen{our} shows that
the axial-vector meson contributes significantly 
to the vector spectral function in hot matter through the mixing:
the presence of the $a_1$ 
reduces the vector spectrum around $M_\rho$
and enhances it around $M_{a_1}$.
The effect of dropping mass of $a_1$ with or without
dropping $\rho$ mass associated with the chiral symmetry
restoration is studied.
It is shown that
the $a_1$-$\rho$-$\pi$ coupling almost 
vanishes at the critical temperature
$T_c$ and thus the V-A mixing becomes very tiny.

In Ref.~\citen{VAmix-dense},
we studied a novel effect of the V-A mixing 
through a set of $\omega\rho a_1$-type interactions
at finite baryon density.
We showed that 
the mixing produces
a clear enhancement of the vector spectral function 
below $\sqrt{s}=m_\rho$,
and that
the vector spectral function 
is broadened due to the mixing.

It is an interesting issue to address a change of the vector 
correlator with the V-A mixing at finite
baryon density in section~\ref{sec:dense}
toward chiral symmetry restoration. 
The mixing (\ref{vaterm}) is chirally symmetric and thus 
does not vanish at the chiral restoration
in contrast to the vanishing V-A mixing near the critical 
temperature $T_c$ without baryon density in section~\ref{sec:hot}.
A spontaneous breaking of Lorentz invariance via the omega 
condensation could increase the mixing strength $C$ near chiral 
restoration~\cite{isb}.
Furthermore,
if meson masses drop due to partial restoration of chiral
symmetry assuming a second- or weak first-order transition
in high baryon density but low temperature region,
the ground state near the critical point may favor
vector condensation even for a moderate mixing strength.
This will be reported elsewhere.

\section*{Acknowledgment}

The work of M.H. is supported in part by
the JSPS Grant-in-Aid for Scientific Research (c) 20540262,
Grant-in-Aid for Scientific
Research on Innovative Areas (No. 2104) 
``Quest on New Hadrons with Variety of Flavors''
from MEXT
and the Global COE Program of Nagoya University 
``Quest for Fundamental Principles in the Universe (QFPU)" 
from JSPS and MEXT of Japan.
The work of C.S. is supported in part 
by the DFG cluster of excellence 
``Origin and Structure of the Universe''.


\bibliographystyle{ws-procs975x65}
\bibliography{ws-pro-sample}

\end{document}